\documentclass[%
 reprint,
 amsmath,amssymb,
prl,
]{revtex4-2}

\usepackage{graphicx}
\usepackage{dcolumn}
\usepackage{bm}
\usepackage{comment}
\usepackage{mathtools}
\usepackage{xcolor, soul}
\usepackage{multirow}
\usepackage{tabularray}
\usepackage{babel}
\renewcommand{\selectlanguage}[1]{}
\usepackage{epstopdf}

\begin{document}


\title{Parametric Frequency Divider based Ising Machines}

\author{Nicolas Casilli$^1$}
\email{N.C. and T.K. contributed equally to this publication.}
\author{Tahmid Kaisar$^2$}
\email{N.C. and T.K. contributed equally to this publication.}
\author{Luca Colombo$^1$}
\author{Siddhartha Ghosh$^1$}%
\author{Philip X.-L. Feng$^2$}
\email{philip.feng@ufl.edu}
\author{Cristian Cassella$^1$}
\email{c.cassella@northeastern.edu\\}
\affiliation{%
$^1$Department of Electrical and Computer Engineering, Northeastern University, Boston, MA 02115, USA
}%
\affiliation{
 $^2$Department of Electrical and Computer Engineering, University of Florida, Gainesville, FL 32611, USA
}%
\date{\today}

\begin{abstract}
We report on a new class of Ising machines (IMs) that rely on coupled parametric frequency dividers (PFDs) as macroscopic artificial spins. Unlike the IM counterparts based on subharmonic-injection locking (SHIL), PFD IMs do not require strong injected continuous-wave signals or applied DC voltages. Therefore, they show a significantly lower power consumption per spin compared to SHIL-based IMs, making it feasible to accurately solve large-scale combinatorial optimization problems (COPs) that are hard or even impossible to solve by using the current von Neumann computing architectures.  
Furthermore, using high quality (\textit{Q}) factor resonators in the PFD design makes PFD IMs able to exhibit a nanoWatt-level power-per-spin. Also, it remarkably allows a speed-up of the phase synchronization among the PFDs, resulting in shorter time-to-solution and lower energy-to-solution despite the resonators' longer relaxation time. As a proof of concept, a 4-node PFD IM has been demonstrated. This IM correctly solves a set of Max-Cut problems while consuming just 600 nanoWatts per-spin. This power consumption is two orders of magnitude lower than the power-per-spin of state-of-the-art SHIL-based IMs operating at the same frequency.

\end{abstract}

\maketitle



Owing to the well-known von Neumann bottleneck \cite{edwards_eager_2016}, most current computing architectures provide limited capability to efficiently solve large-scale nondeterministic polynomial-time (NP) hard problems within a reasonable amount of time \cite{peper_end_2017}. 
To address this limitation, a new approach to solving NP-hard problems has emerged in the form of hardware solvers called Ising machines (IMs). An IM can be defined as a network of \emph{artificial} spins \cite{bohm_A_2019}, arranged and interconnected according to the problem at hand. This machine can accurately solve a combinatorial optimization problem (COP) by identifying the spin-state configuration that minimizes the corresponding Ising Hamiltonian \cite{mohseni_ising_2022,lucas_ising_2014,kalinin_computational_2022}. 
Several systems have been developed in recent years to perform an efficient minimization of the Ising Hamiltonian, including:  D-Wave systems \cite{harris_experimental_2010,johnson_quantum_2011,afoakwa_cmos_2020}, Coherent Ising machines (CIMs) \cite{inagaki_coherent_2016,marandi_network_2014,cen_large-scale_2022}, {photonic IMs} \cite{pierangeli_large_2019,prabhu_accelerating_2020,yamashita_low-rank_2023}, SRAM-based IMs \cite{su_flexspin_2022,yamaoka_overview_2019}, GPU-based IMs \cite{cook_gpu_2019},  and oscillator-based Ising machines (OIMs) \cite{chou_analog_2019,moy_1968-node_2022,wang_oim_2019,wang_oscillator-based_2017,csaba_coupled_2020, bybee_efficient_2022,ochs_an_2021, ahmed_probabilistic_2021,zhang_oscillator-network_2022}. D-Wave systems rely on superconducting devices \cite{harris_experimental_2010} requiring cryogenic operating temperatures near zero kelvin to function properly. Consequently, they are bulky and consume a considerable amount of power due to the necessity of cryogenic refrigeration \cite{johnson_quantum_2011}. CIMs utilize fiber-based optical parametric oscillators  \cite{marandi_network_2014,cen_large-scale_2022} to generate the spins and field-programmable-gate-arrays (FPGAs) to digitize the spins' coupling \cite{marandi_network_2014,inagaki_coherent_2016}. As a result, they are also hardly usable when targeting a small form-factor and a low power consumption. {Alternative photonic IMs based on spatial light modulation} \cite{pierangeli_large_2019} {or recurrent Ising sampling} \cite{prabhu_accelerating_2020} {have also been reported, showing promise for solving large-scale COPs. However, relying on these solvers also comes with challenges, {primarily related to unfavorable times-to-solution caused by the required intense signal processing} \cite{yamashita_low-rank_2023}. {SRAM-based and GPU-based IMs are digital hardware implementations of the simulated annealing algorithm or of one of its variants} \cite{wang_oscillator-based_2017,mohseni_ising_2022}. These IMs can be manufactured using the same complementary metal-oxide-semiconductor (CMOS) fabrication processes \cite{afoakwa_cmos_2020} utilized for mass production of the integrated circuits in consumer electronics, offering significant benefits in terms of production cost, reprogrammability, and form-factor. {However, the performance of these IMs depends on the problem being solved and can be significantly degraded for problems requiring heavy sequential computation}} \cite{su_flexspin_2022,yamaoka_overview_2019,hamerly_experimental_2019,wang_oim_2019,cook_gpu_2019}. 
\begin{figure}[t]
\includegraphics[width=\linewidth]{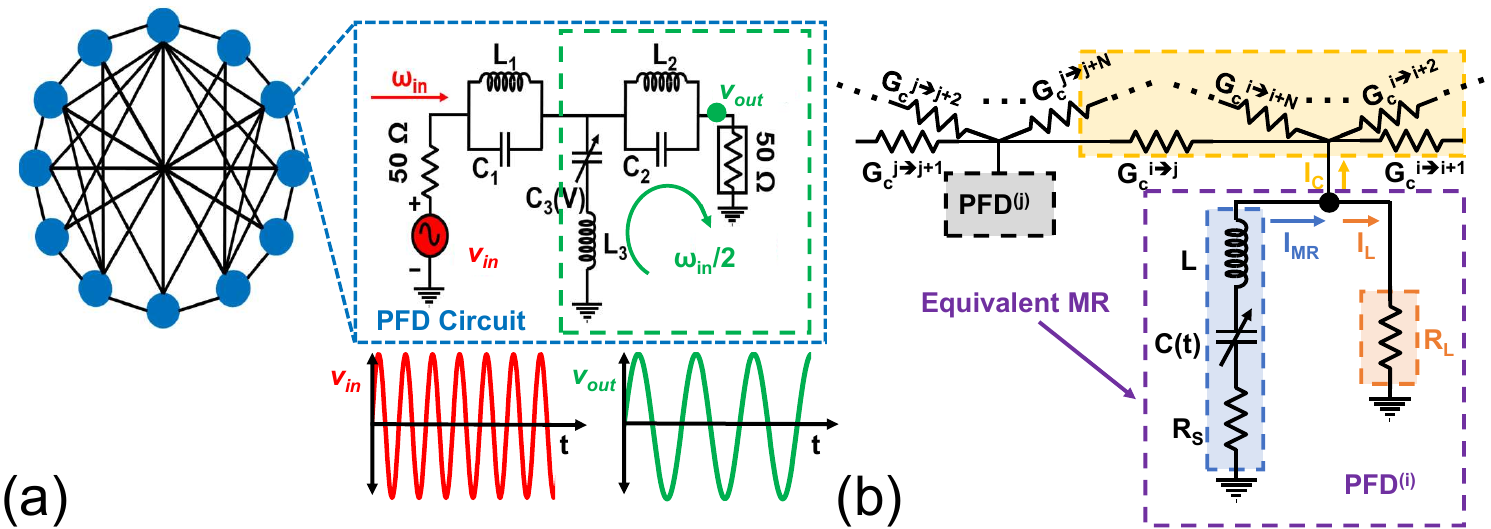}
\caption{\label{fig:epsart}a) Schematic view of a PFD IM; b) Schematic view of a network of coupled PFDs, where each PFD is described by an electrical Mathieu resonator (MR). }
\end{figure}
For these reasons, the pursuit of highly miniaturized and low-power IMs has recently shifted towards OIMs, {whose physics-inspired processing enables a higher degree of parallelism during the computation compared to digital solvers} \cite{yamaoka_overview_2019,hamerly_experimental_2019}. 

OIMs leverage the collective dynamics of networks of bistable coupled electronic oscillators to perform the computation in an analog fashion. {Among the demonstrated OIMs} \cite{mohseni_ising_2022}{, those using ``parametrons" as spins were the first ones to be proposed} \cite{calvanese_strinati_can_2021,heugel_ising_2022,english_an_2022,bello_persistent_2019}. {Parametrons attain phase bistability by triggering a parametric oscillation in a circuit composed of one nonlinear resonator. In this regard, the dynamics of coupled parametric oscillators have been studied in the last few years to benchmark the computing performance achievable by CIMs and by parametron based IMs} \cite{bello_persistent_2019,calvanese_strinati_coherent_2020,calvanese_strinati_can_2021}. {Yet, a full investigation of the accuracy of the retrieved problem solution when using an annealing schedule or when relying on resonant devices with high quality factor ($Q$), like the micro and nano photonic and electromechanical resonators available today} \cite{feng_concepts_2022,pillai_piezoelectric_2021,bereyhi_perimeter_2022,cassella_aluminum_2016}{, is currently missing. Moreover, all the reported parametrons require hundreds of milliWatts of input power to activate their oscillation. Such a high power consumption motivates why no attempt has ever been made to build large-scale electronic IMs based on parametrons (see Supplemental Material Section S3.3 for more information about the parametrons developed to date \cite{Casilli_SM}).} On the other hand, OIMs utilizing subharmonic-injection-locked (SHIL) oscillators as spins have garnered significant attention in recent years \cite{chou_analog_2019, moy_1968-node_2022, wang_oim_2019, wang_oscillator-based_2017, csaba_coupled_2020,bybee_efficient_2022,ochs_an_2021}. In these OIMs, dubbed here as ``SHIL IMs", an artificial spin is represented by the bistable phase of a SHIL oscillator's output signal, which can be shifted by either $0$ or $\pi$ with respect to the output phase of a reference oscillator. {SHIL IMs are generally analyzed by using the Kuramoto model} \cite{acebron_the_2005}{, which only considers the phase of the SHIL oscillators' output signal and not the amplitude.} The power consumed by each oscillator in SHIL IMs is typically in the hundreds of $\mu$Watts-range due to the need to sustain the oscillation, trigger the injection-locking regime, and synthesize the spin-coupling \cite{wang_oscillator-based_2017,mohseni_ising_2022}. As a result, the current SHIL IMs are also not easily scalable to solve realistically sized NP-hard problems while maintaining a low power consumption \cite{chou_analog_2019,moy_1968-node_2022, mohseni_ising_2022}.

In this Letter, we present a class of OIMs referred to as parametric frequency divider based IMs (PFD IMs). 
In recent years, PFDs have been used for sensing \cite{hussein_chip-less_2021, hussein_capturing_2021}, signal processing \cite{hussein_reflective_2021,ramirez_nonlinear_2008}, and frequency generation \cite{villanueva_nanoscale_2011,cassella_phase_2017}. Like the previously reported parametrons, PFDs rely on a nonlinear reactance, such as a diode or a varactor, to passively activate a parametric oscillation at half of their driving signal's frequency ($\omega_{0}$) when the input power levels exceed a certain threshold (\emph{$P_{th}$}). {Yet, in order to do so, they couple a set of four harmonically related resonances to boost the effectiveness of the parametric modulation in their circuit, thereby enabling \emph{$P_{th}$} values that are orders of magnitude lower than previously demonstrated for parametrons} \cite{hussein_systematic_2020,casilli_nonvolatile_2023}.  

As depicted in Fig. 1-a, a PFD can be characterized as a two-port electrical network formed by two circuit meshes interconnected through a shunt branch that contains the nonlinear capacitor. The input mesh is driven by the PFD's input signal [\emph{$v_{in}(t)$}], which modulates the capacitance of the nonlinear capacitor at an angular frequency $\omega_{in}$. Each mesh incorporates a set of notch filters. These filters constrain $v_{in}(t)$ and the output signal, $v_{out}(t)$, within the PFD's input and output meshes respectively, allowing to analyze the PFD's behavior at each frequency by examining just one mesh. As described in \cite{hussein_systematic_2020}, the reactive components in the output mesh of a PFD are selected to series-resonate at half of the input natural frequency (e.g., $\omega_{in}$=2$\omega_{0}$) when neglecting the capacitance modulation induced by $v_{in}(t)$. This permits a mapping of the PFD's operation at or near $\omega_{0}$ with only one second-order differential equation describing the voltage across the nonlinear capacitor. This mapping is equivalent to an electrical realization of a Mathieu resonator (\emph{MR}, see Fig. 1-b) \cite{rhoads_generalized_2006}. Such an \emph{MR} has a \emph{Q} equal to 1/(2$\gamma_{tot}$), where $\gamma_{tot}$ models the resonator's damping (e.g., $\gamma_{tot}$=$\omega_{0}$$C_{av}$$R_{tot}$/2, where $C_{av}$ is the average capacitance of the nonlinear capacitor for $v_{in}(t)$=0). $R_{tot}$ is equal to $R_{L}$+$R_s$, where $R_s$ denotes the \emph{intrinsic} losses in the resonant system (e.g., the total resistance in the PFD's output mesh, $R$, combined with the resistance, $R_d$, capturing the Ohmic losses in the nonlinear capacitor's electrodes and dielectric film). Also, the \emph{MR} has a resonance angular frequency in the absence of modulation equal to $\omega_{0}$, and this frequency is periodically varied at a rate equal to $\omega_{in}$. 
In this regard, we denote the magnitude of the resonance frequency modulation caused by the input signal at $\omega_{in}$ as $p$. 
As in its mechanical counterpart, the \emph{MR} describing the operation of a single PFD enters a period-doubling regime for $p$-values larger than a certain threshold ($p_{th}$) equal to 4$\gamma_{tot}$. More information on the \emph{MR}-model of a PFD is provided in {Supplemental Material Section S1 \cite{Casilli_SM}}.  

In order to demonstrate that networks of PFDs can be used as IMs, we analyze their interacting dynamics when they are coupled. This can be done by considering a number ($N$) of \emph{MRs} with the same \emph{Q} and $\omega_{0}$ values, and we assume all couplings among the \emph{MR}s to be purely dissipative (e.g., the PFDs are coupled through resistors connected to their output meshes). To this end, small coupling conductances ($\epsilon$\textit{$G_{i,j}$}) with generic indices \emph{i} and \emph{j} can be used to map the interaction between the generic \emph{i}-th and \emph{j}-th \emph{MRs}, as shown in Fig. 1-b. In particular, a summation can be used to capture all the interactions that any given \emph{MR} is subject to based on the targeted problem to solve  (see Supplemental Material Section S2 \cite{Casilli_SM}). As an example, we report in Eq. (1) the \emph{MR}-equation we have used to analyze the dynamics of the \emph{i}-th \emph{MR} during our analytical treatment. The variables $\textit{v}_{i,j}$ in Eq. (1) describe the voltage across the nonlinear capacitors in the \emph{i}-th and \emph{j}-th \emph{MRs}, respectively.    
\begin{equation}
\begin{split}
\MoveEqLeft
       {\ddot{v_i}}+2\epsilon\gamma_{tot} \omega_{0} \dot{v_i}+ \omega_{out}^2 {v_i}+ \\
       &2\gamma_{L} \omega_{0}{R_{L}}{\sum\limits_{j \neq i} \epsilon{G_{ij}}}{\dot{v_j}}=0.
\end{split}
\end{equation}

Differently from the equation of motion of only one PFD (Supplemental Material Section S1 \cite{Casilli_SM}), Eq. (1) includes an additional damping parameter, $\gamma_L$, equal to $\omega_{0}C_{av}R_L$/2. Also, the  angular resonance frequency for all \emph{MR}s incorporates a ``pump-depletion" term proportional to $\beta$ [Eq. (2)] that is responsible for the saturation of the voltage across their nonlinear capacitors for {$p$}$>${$p_{th}$}.
\begin{equation}
\begin{split}
\MoveEqLeft
       \omega_{out}^2=\omega_{0}^2[1+\epsilon p(1-\beta (v_i)^2)\sin{({2\omega_{0}}t)}].
\end{split}
\end{equation}

It is important to point out that in Eqs. (1)-(2) both $p$ and $\gamma_{tot}$ are assumed to be small and are consequently scaled by a small parameter $\epsilon$. From Eqs. (1)-(2), we can apply the Multiple Scales Method \cite{kumar_numerical_2020,calvanese_strinati_theory_2019} to derive a system of first-order differential equations [Eq. (3)] governing the dynamics of the complex voltage amplitude for the slow time scale $\tau$=$\epsilon$$t$. For the derivation of Eq. (3), we have assumed the lowest order response of $v_i$ to be expressible as $B_i(\tau)$$e^{i\omega_{0}t}$+$B_i^*(\tau)$$e^{-i\omega_{0}t}$, where $B_i^*(\tau)$ is the complex conjugate of $B_i(\tau)$. Also, when \emph{MR}s are used to solve a COP, we expect the solution to be encoded in the phase [$\phi$($\tau$)] of the complex amplitude reached at steady state by all the adopted \emph{MR}s, similarly to what happens in CIMs and SHIL IMs. Therefore, from the real [$B_{i,re}$($\tau$)] and imaginary [$B_{i,im}$($\tau$)] parts of $B_i$($\tau$) we can calculate $\phi_i$($\tau$) as $\arctan$[$B_{i,im}$($\tau$)/$B_{i,re}$($\tau$)]. We then evaluate the steady-state value ($\Phi_i$) of $\phi_i$($\tau$), and the same procedure is run for all the adopted \emph{MR}s. Independently of the problem that needs to be solved, each \emph{MR} can only reach two $\Phi$-values, namely $0$ or $\pi$, giving each PFD in a PFD IM the ability to passively emulate the dynamics of an Ising spin. {In this regard, similar to CIMs, that utilize parametric dynamics to achieve phase bistability in the optical domain, the ground state solution identified by a PFD IM is governed by the minimization of a Lyapunov function considering both amplitude and phase dynamics} \cite{wang_bifurcation_2023,roychowdhury_global_2022}. {This computational principle is also similar to that of SHIL IMs, with the key distinction that the Lyapunov function governing SHIL IMs considers only the system's phase dynamics} \cite{wang_oscillator-based_2017,roychowdhury_global_2022}. More information about the dynamics of coupled PFDs are provided in Supplemental Material Section S4 \cite{Casilli_SM}. 

\begin{equation}
\begin{aligned}
\MoveEqLeft
    B_i'(\tau)=\frac{1}{4}{[(p\omega_0\beta)(B_i^3-}3B_iB_i^{*^2}) + p\omega_0B_i^*-\\
       &4\omega_{0}{B_i}\gamma_{tot}-4\omega_{0}R_{L}\gamma_{L}{\sum\limits_{j \neq i} {G_{ij}}}{B_j}].
\end{aligned}
\end{equation}

Starting from Eq. (3) (see Supplemental Material Section S2 \cite{Casilli_SM}), we can evaluate the performance of a PFD IM when computing the solution of a COP over $N$ variables, with each variable mapped to a specific PFD. In this regard, the performance of IMs are assessed based on several factors, including the probability of achieving a spin configuration that matches or closely matches the problem solution [i.e., the ``probability-of-success" ($P$)], the time required to obtain a solution [i.e., the ``time-to-solution" ($T_S$)], the power consumption of each spin [i.e., the ``power-per-spin" ($PW_{spin}$)], and the energy consumed by the entire machine during the computation [i.e., the ``energy-to-solution" ($E_S$)]. In order to evaluate these computing performance metrics for PFD IMs, we construct a specific coupling matrix [$G$] for each problem, with dimension $N$$\times$$N$. Each row of [$G$] incorporates the conductance used to couple one specific PFD to any other PFDs. For each PFD, the phase of the slow complex amplitude [Eq. (3)] of its corresponding equivalent \emph{MR} equation is numerically computed to obtain $\Phi$. 
Then, $P$ is determined based on a desired accuracy level ($A$). $A$ is the minimum tolerated accuracy for the problem solution and its value ranges from 0 to $100\%$. Depending on whether we are looking at the probability to reach the ground-state (e.g., the global minimum for the targeted COP that identifies a 100$\%$ accurate solution) or at the probability to reach a close enough solution to the ground-state, with an accuracy higher than $A$ but lower than 100$\%$, $P$ can be re-written as \emph{$P_{GS}$} or \emph{$P_{A}$}, respectively. Both \emph{$P_{GS}$} and \emph{$P_{A}$} can be computed for any targeted problem by solving it multiple times. After determining $P_A$, $T_S$ can be calculated as \cite{mohseni_ising_2022}:
\begin{equation}
\begin{split}
{T_{S}}={\tau_{\phi}}\times[log_{(1-P_A)}(A)], 
\end{split}
\end{equation}
where $\tau_{\phi}$ is the time that it takes on average for the phases of the slow-complex amplitudes of all coupled \emph{MR}s to reach their final value when multiple problem-runs are executed. 
It is worth mentioning that the achievement of optimal computing performance can pass through the adoption of an annealing step, similarly to prior SHIL IMs \cite{chou_analog_2019}. To this end, $p$ is gradually increased up to 1.005$p_{th}$ from an initial value equal to 0.995$p_{th}$ following an exponential trend [e.g., $p(t)$ = $p_{th}$(0.995+0.01(1-$e^{-t/\tau_{ann}}$)), where $\tau$$_{ann}$ is the annealing rate and $t$ is the time].  
After determining $T_{S}$, bearing in mind that $p$ reflects the voltage magnitude at $\omega_{in}$ across the varactor in each PFD and that $P_{th}$ is proportional to $p_{th}^2$, we can estimate $E_S$ for any problem as $N$$P_{th}$\(\int_{0}^{T_{S}} (0.995+0.01(1-e^{-t/\tau_{ann}}))^2\,dt\) \cite{chou_analog_2019,wang_oim_2019}. The $P_{th}$ value considered during the computation of $E_S$ can be directly found through a circuit simulation of a PFD ({Supplemental Material Section S3} \cite{Casilli_SM}). It is also worth mentioning that the ability to passively generate Ising dynamics without active components allows us to consider the Johnson noise generated by the \emph{MRs}' resistors as the only noise process affecting the \emph{MRs}' circuit \cite{pierret_advanced_2003}. 

\begin{figure}[b]
\centering
\includegraphics[width=\linewidth]{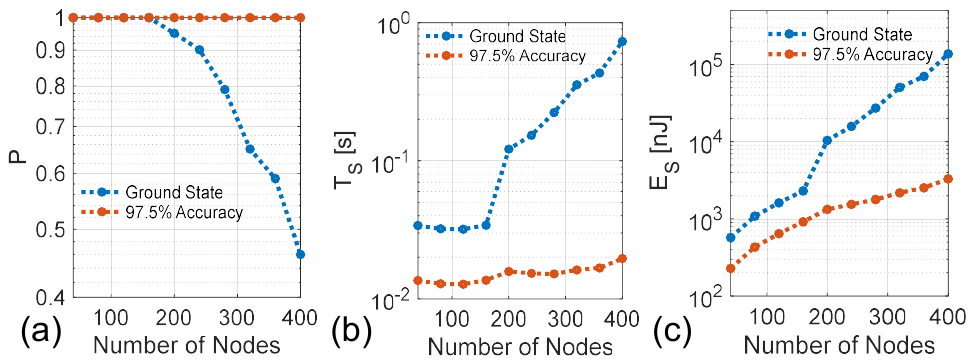}
\caption{\label{fig:wide} Numerically computed trends of a)  $P_{GS}$ and $P_{97.5\%}$, b) $T_{S}$ for $A$=97.5$\%$ or $A$=100$\%$, and c) $E_S$ for $A$=97.5$\%$ or $A$=100$\%$  {vs. increasing $N$ in M{\"o}bius ladder problems when $\tau_{ann}$ = 1 sec}. }
\end{figure}

To analyze and benchmark the performance of our PFD IMs, we choose to connect all the PFDs in a M\"{o}bius ladder configuration and to solve a set of unweighted Max-Cut problems of varying sizes \cite{harris_experimental_2010,chou_analog_2019,mohseni_ising_2022}. {COPs with a M\"{o}bius ladder graph are considered low-complexity sparse problems} \cite{strinati_all-optical_2021}. {Consequently}, their correct solution can be numerically calculated, allowing to easily verify whether the solution found by a PFD IM is correct and, if it is not correct, to evaluate the difference in the computed number of cuts with respect to the expected value \cite{kalinin_computational_2022}. The number of cuts computed by a PFD IM is equivalent to the total number of paths of the problem-graph connecting PFDs with different output phases \cite{kalinin_computational_2022,ochs_an_2021}.

By relying on our analytical model, we first investigated $P_A$, $T_S$ and $E_S$ when scaling $N$ in the graph from 40 to 400 and when assuming specific values of tolerated accuracy ($A$=100$\%$ and $A$=97.5$\%$) frequently used for benchmarking IMs \cite{hamerly_experimental_2019}. During this study, we initially considered a $Q$=50, which is approximately the same $Q$ of the resonators used by the PFDs assembled in this work. Also, we assumed an $\omega_0$=2$\pi$$\times10^6$ rad/sec, which coincides with the output angular frequency of our assembled PFD IM. For each considered $N$ value, we computed the problem solution 100 times. 
This allowed us to determine $P_{GS}$ and $P_{97.5\%}$ (see Fig. 2-a), together with the number of cuts identified by each executed problem-run. We found that the likelihood of generating a 100$\%$ accurate solution rapidly decays with respect to $N$, which is in line with what is generally observed in other IMs \cite{chou_analog_2019,wang_oim_2019,afoakwa_cmos_2020,calvanese_strinati_can_2021,yamaoka_overview_2019, mohseni_ising_2022}. Nevertheless, PFD IMs retain a $100\%$ likelihood of calculating a cut-size within $2.5\%$ of the highest possible number of cuts. In addition, after identifying $\tau_{\phi}$, we computed $T_S$ and $E_S$ vs. $N$ when assuming a 100$\%$ or a 97.5$\%$ accuracy (see Figs. 2-b,c). In this regard, during the calculation of $E_S$ we assumed a $P_{th}$ value (600 nW) matching what we simulated and measured in our experiments. Evidently, we found an $E_S$ value of {135 $\mu$J (3.3 $\mu$J)} when assuming a 100$\%$ (97.5$\%$) minimum tolerated accuracy in the calculation of $T_{S}$.  

Subsequently, we conducted a second study driven by the growing accessibility of high-$Q$ chip-scale resonator technologies that can be manufactured using the same semiconductor processes employed for solid-state varactors and diodes \cite{feng_concepts_2022,pillai_piezoelectric_2021,bereyhi_perimeter_2022,cassella_aluminum_2016,kaisar_five_2023}. 
\begin{figure}[t]
\includegraphics[width=\linewidth]{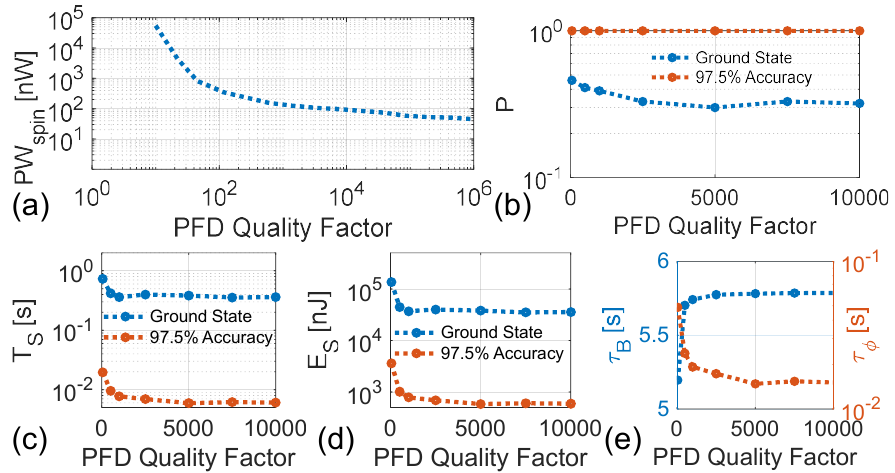}
\caption{\label{fig:epsart} Numerically computed trends for $\tau_{ann}$ = 1 sec vs. $Q$ of a) $PW_{spin}$, b) $P_{GS}$ and $P_{97.5\%}$, c) $T_S$ when considering $A$=97.5$\%$ or $A$=100$\%$, d) $E_S$ when considering $A$=97.5$\%$ or $A$=100$\%$, and e) $\tau_B$ and $\tau_{\phi}$.}
\end{figure}
In particular, it is reasonable to question whether incorporating these resonators in place of the $L-C$ resonators currently used to construct PFDs could enhance the performance of PFD IMs. 
Therefore, we analyzed the performance of PFD IMs vs. $Q$. First, we calculated the trend (Fig. 3-a) of $PW_{spin}$ vs. $Q$ through a circuit simulator ({see Supplemental Material Sections S3 and S5} \cite{Casilli_SM}). Interestingly, we found that relying on resonators with $Q$s higher than $10^6$ permits a reduction of  $PW_{spin}$ down to 60 nW, which is three orders of magnitude lower than the power required by each oscillator in state-of-the-art SHIL IMs. It is worth emphasizing that the saturation of $PW_{spin}$ for $Q$ values higher than $10^6$ originates from the fact that $R_{L}$ and $R_{d}$ do not scale down with $Q$. We also analyzed $P$ (Fig. 3-b), $T_S$ (Fig. 3-c) and $E_S$ (Fig. 3-d) vs. $Q$ for a 400-node PFD IM solving the same Max-Cut problem we considered in Fig. 2 when assuming minimum tolerated accuracy levels of 100$\%$ and 97.5$\%$, as in our first study. Interestingly, we found that relying on higher $Q$ resonators does not change significantly the values of \emph{$P_{GS}$} and \emph{$P_{97.5\%}$} with respect to the values found for a $Q$ equal to 50 in Fig. 2. However, $T_S$ reduces when assuming higher $Q$ values, despite the fact that high-$Q$ resonators inherently exhibit a longer relaxation time. This is due to the fact that $\tau_{\phi}$ shortens when considering high $Q$ values, even though a longer relaxation time ($\tau_B$) is needed for the \emph{MR}s to reach their steady-state amplitude (Fig. 3-e). Consequently, PFD IMs that rely on higher $Q$ resonators inherently exhibit a lower $E_S$ than their lower $Q$ counterparts. As such, they are better suited for addressing COPs with a large number of variables.
Finally, the impact of $\tau_{ann}$ on the computing performance of PFD IMs has also been analyzed for different $N$ and $Q$ values. We found that using a slower annealing rate {when tackling large M{\"o}bius ladder problems} remarkably leads to lower $T_S$ and $E_S$ values, despite the increase of $\tau_{\phi}$ and independently of the \emph{MRs}' \emph{Q} value. {We verified (Supplemental Material Section S5 \cite{Casilli_SM}) that this improved performance can be attributed to a significant reduction in amplitude heterogeneity for longer annealing rates} \cite{albash_analog_2019,leleu_destabilization_2019}.  

 \begin{figure}[t]
\centering
\includegraphics[width=\linewidth]{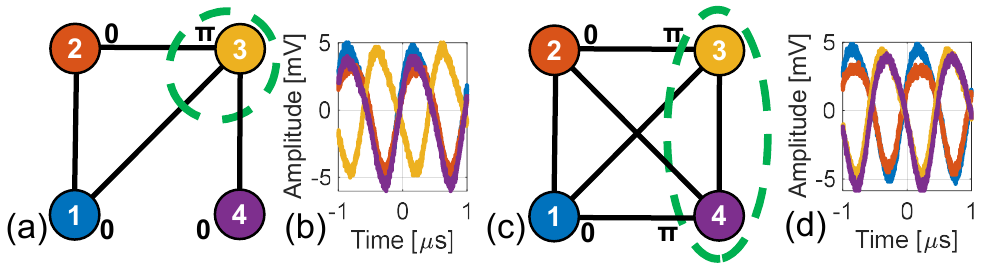}
\caption{\label{fig:wide}a-d) Graphs and PFDs' output voltage waveforms relative to two Max-Cut problems tackled with the PFD IM built in this work. }
\end{figure}

As a proof of concept, we built the first prototype of a PFD IM and we employed it to solve different unweighted Max-Cut problems, as in \cite{chou_analog_2019}. Four identical PFDs designed to work with a $\omega_{in}$ value of 4$\pi$*10$^6$ rad/s were assembled on a printed-circuit-board (PCB) by using off-the-shelf inductors and capacitors to create resonant tanks with a $Q$ value of nearly 50. Identical 2$k\Omega$ coupling resistors, corresponding to 500 $\mu S$ coupling conductances (Fig. 1-b), were used to couple the four PFDs according to the specific problem to solve. By running an electrical characterization of our PFDs, we were able to experimentally demonstrate a $PW_{spin}$ value of 600 nW, which is the lowest one ever recorded for OIMs. It is worth emphasizing that all PFDs were designed to generate their subharmonic oscillation without requiring DC-voltages, thus without consuming any DC power for biasing the circuit. 
Fig. 4 shows the graphs of two of the nine Max-Cut problems investigated and solved by the PFD IM built in our experiments, together with the corresponding measured PFDs' output voltage. Evidently, the computed phase distribution matches the expected correct solution \cite{chou_analog_2019} for every problem we evaluated.  
A description of the experimental setup used during the testing of the assembled PFD IM is provided in Supplemental Material Section S6 \cite{Casilli_SM}, together with the graphs and output voltage waveforms for the other Max-Cut problems we have solved. 


{In conclusion, we have introduced PFD IMs and studied their computing performance when tackling various M{\"o}bius ladder problems with up to 400 nodes. Our findings suggest that incorporating high-$Q$ resonators in the PFDs' design and using an annealing schedule allow to decrease $PW_{spin}$ down to the nanoWatt-range, shorten $T_S$ to less than {0.75 $s$}, boost the $P_{GS}$ up to 46$\%$ and achieve an $E_S$ of {135 $\mu J$} for a 400-node M{\"o}bius ladder problem. We have also designed, built, and tested a prototype of a PFD IM that integrates four PFDs to solve several different Max-Cut problems. This prototype achieves a $PW_{spin}$ of 600 nW by relying on off-the-shelf $L-C$ resonators with a $Q$ near 50, and always retrieves the correct solutions for all the problems we have tackled. The demonstrated $PW_{spin}$ is the lowest one ever reported for OIMs. Further investigation and performance evaluation will be required in the future to characterize the performance of PFD IMs for generic NP instances with densely connected graphs, beyond the Möbius ladder problems discussed in this Letter.} 


\nocite{suarez_analysis_2009}
\nocite{cassella_low_2015}
\nocite{rober_novel_2013}
\nocite{wang_coherent_2013}
\nocite{bohm_order_2021}
\nocite{nifle_new_1992}

\bibliography{TeX/PRL_2023_Paper_updated}

\end{document}



\title{Parametric Frequency Divider based Ising Machines - Supplementary Material}

\author{Nicolas Casilli$^1$}
\author{Tahmid Kaisar$^2$}
\author{Luca Colombo$^1$}
\author{Siddhartha Ghosh$^1$}%
\author{Philip X.-L. Feng$^2$}
\author{Cristian Cassella$^1$}
\affiliation{%
$^1$Department of Electrical and Computer Engineering, Northeastern University, Boston, MA 02115, USA
}%
\affiliation{
 $^2$Department of Electrical and Computer Engineering, University of Florida, Gainesville, FL 32611, USA
}%


\maketitle

\section{\label{sec:level1} Acronyms}

\begin{tabular}{ p{1.4cm}|p{6.7cm}  }
IM  & Ising Machine \\
PFD & Parametric Frequency Divider \\
SHIL & Subharmonic Injection Locked \\
\emph{Q} & Quality Factor \\
NP & Nondeterminisitc Polynomial-time \\
COP & Combinatorial Optimization Problem \\
PFD IM & Parametric Frequency Divider-based Ising Machine \\
SHIL IM & Subharmonic Injection Locking-based Ising Machine \\
CIM & Coherent Ising Machine \\
SRAM & Static Random Access Memory \\
GPU & Graphics Processing Unit \\
OIM & Oscillator Ising Machine \\
FPGA & Field-Programmable Gate-Arrays \\
$\omega_{0}$ & Angular frequency of the output of PFDs \\
$P_{th}$ & Power Threshold \\
$v_{in}(t)$ & Input signal of PFDs \\
$\omega_{in}$ & Angular frequency of $v_{in}(t)$ which modulates the nonlinear capacitor in the PFDs \\
$v_{out}(t)$ & Output signal of PFDs \\
$\omega_{out}$ & Angular frequency of $v_{out}(t)$ \\
$R_L$ & Load resistance of PFDs or Mathieu's Resonators \\
\emph{MR} & Mathieu Resonator \\
$\gamma_{tot}$ & Damping of the \emph{MR} when also considering $R_L$ \\
$C_{av}$ & Capacitance of the nonlinear capacitance for $v_{in}(t) = 0$ \\
$R_{tot}$ & Total resistance of the PFDs or \emph{MR}s \\
$R_s$ & Intrinsic losses in the resonant system of the \emph{MR} or PFD \\ 
$R$ & Total resistance in the output mesh of PFDs \\
$R_d$ & Ohmic losses of the nonlinear capacitor \\
$p$ & The depth of the resonance frequency modulation in $v_{in}(t)$ \\
$p_{th}$ & Threshold of the modulating signal that excites the subharmonic signal \\
$N$ & Number of nodes \\
$\epsilon$ & Small-scale parameter \\
$G_{i,j}$ & Conductance coupling $MR_i$ to $MR_j$ at their outputs\\
$\gamma_L$ & Damping of the \emph{MR} due to losses in the load \\
$\tau$ & Slow time scale \\
$v_i$ & Voltage across the nonlinear capacitor \\

\end{tabular}

\begin{tabular} { p{1.4cm}|p{6.7cm}  }
$B_i(\tau)$ & Slow-varying complex amplitude term of $v_i$ \\
$\phi(\tau)$ & Phase of the complex amplitude of the \emph{MR}s \\
$\Phi_i(\tau)$ & Phase of the complex amplitude of the i-th \emph{MR} taken at the steady state \\
$P$ & Probability-of-Success \\
$T_S$ & Time-to-solution \\
$PW_{spin}$ & Power-per-spin \\
$E_S$ & Energy-to-solution \\
$G$ & Coupling matrix \\
$A$ & Desired solution accuracy for $P$ \\
$GS$ & Ground-state \\
$P_{GS}$ & Probability-of-success for achieving GS \\
$P_A$ & Probability-of-success for achieving a solution that is within (1-A)\% of the GS \\
$\tau_{\phi}$ & Time that it takes, on average, for the phases of the slow-varying complex amplitudes to reach their final value \\
{$\tau_B$} & {Relaxation time needed for the MRs to reach their steady-state amplitude values} \\
$\tau_{ann}$ & Annealing rate \\
t & Elapsed time during the runs of the IM \\
\emph{pAG} & Power Auxiliary Generator \\
$R_{AG}$ & Internal resistance of \emph{pAG} \\
{$CV$} & {Coefficient of Variation, describing the ratio between the standard deviation and the mean of a set} \\
{KVL} & {Kirchhoff's Voltage Law} \\
{KCL} & {Kirchhoff's Current Law}
\end{tabular}

\clearpage
\newpage
\clearpage
\newpage

\pagebreak

\setcounter{equation}{0}
\renewcommand{\theequation}{S\arabic{equation}}
\setcounter{figure}{0}
\renewcommand{\thefigure}{S\arabic{figure}}

\newpage
\section{\label{sec:level1} Section S1}
\begin{center}
    \textbf{Modelling PFDs as Mathieu's Resonators}
\end{center}

Starting from the PFD's electrical circuit shown in Fig. 1-a and as discussed in the main manuscript, the PFD's behavior at or near $\omega_0$ can be described by a single resonant branch capturing the behavior of the PFD's output mesh around $\omega_0$. This resonant branch is formed by the series of $L$, $C(t)$, and $R_{tot}$, and it conforms to the structure of an electrical \emph{MR} \cite{rhoads_generalized_2006} with total damping proportional to $R_{tot}$. We can analyze the behavior of this circuit by studying its corresponding {Kirchhoff's Voltage Law} (KVL) equation in terms of the voltage ($v$) across the \emph{MR}'s nonlinear capacitor [see Eq. (S1)]:

\begin{equation}
\begin{split}
\MoveEqLeft
    {L}{C(t)}{\ddot{v}}+ v+\epsilon{R_{tot}} {C_{av}} \dot{v} =0.
\end{split}
\end{equation}


In Eq. (S1), $C_{av}$ maps the average capacitance of the PFD's nonlinear capacitor during the period of the input signal ($T_{in}$ = 2$\pi$/$\omega_{in}$). Also, a small expansion parameter ($\epsilon$) is used to scale the value of $R_{tot}$. In this regard, Eq. (S1) has been truncated at the first order of $\epsilon$ (e.g., $\epsilon^2$=0, $\epsilon^3$=0, etc.). Bearing in mind that ${\omega_{0}}^2$=$1/L{C_{av}}$, Eq. (S1) can be rewritten as: 

\begin{equation}
\begin{split}
\MoveEqLeft
       {\ddot{v}}+ \omega_{out}^2 \epsilon{R_{tot}} {C_{av}} \dot{v}+ \omega_{out}^2 {v}=0.
\end{split}
\end{equation}

By expressing ${\omega_{out}}^2$ as shown in Eq. (2), defining $\gamma_{tot}$ as $\omega_{out}$$C_{av}$$R_{tot}$/2, and once again disregarding terms proportional to higher powers of $\epsilon$, we can represent Eq. (S2) as follows: 

\begin{equation}
\begin{split}
\MoveEqLeft
      {\ddot{v}}+\omega_{0}\epsilon2\gamma_{tot}  \dot{v}+ \\
      &\omega_{0}^2[1+\epsilon p(1-\beta (v)^2)\sin{({2\omega_{0}}t)}]{v}=0.
\end{split}
\end{equation}

We can now apply the Multiple Scale Method (MSM) \cite{kumar_numerical_2020}. In order to do so, we start by separating the fast- and slow-changing timescales of $v$ as $v_{}$=${v_{}}^{(0)}$+$\epsilon$${v_{}}^{(1)}$, where ${v_{}}^{(0)}$ and  ${v_{}}^{(1)}$ are the zeroth-order and first-order terms of $v_{}$, respectively. Also, we rewrite the time derivatives as $d/dt$=$D_{0}$+$\epsilon$$D_{1}$, and $d^2/dt^2$=${D_{0}}^2$+2$\epsilon$$D_{0}$$D_{1}$, where $D_{0}$=$\partial$/$\partial t$ and $D_{1}$=$\partial$/$\partial \tau$. This allows to rewrite Eq. (S3) as:
\begin{equation}
\begin{split}
\MoveEqLeft
      D_0^2 v^{(0)}+2\epsilon {D_0}{D_1}{v}^{(0)}+\epsilon D_0^2 v^{(1)}+\\
      &2\epsilon \omega_{0}\gamma_{tot}[{D_0}v^{(0)}+\epsilon D_0 v^{(1)} + \epsilon D_1 v^{(0)}]+\\
      &\omega_0^2v^{(0)} + {\omega_{0}}^2 \epsilon v^{(1)}+ \\
      &\omega_0^2 \epsilon p (1-\beta {v}^2) sin (2\omega_0 t)  v^{(0)}=0.
\end{split}
\end{equation}

By retaining only the terms proportional to $\epsilon$, we can write Eq. (S4) as:
\begin{equation}
\begin{split}
\MoveEqLeft
      D_0^2v^{(1)} + \omega_0^2v^{(1)}  + 2D_0D_1v^{(0)} + \\
      & 2\omega_0\gamma_{tot}D_0v^{(0)} + \\
      &\omega_{0}^2 p(1-\beta {v}^2) sin (2\omega t)  {v}^{(0)}=0.
\end{split}
\end{equation}

Given that we expect the zeroth-order of the voltage across the \emph{MR}'s capacitor to vary at a rate equal to $\omega_0$, we can rewrite the lowest order response of $v$ as $v^{(0)}$=$B_{}(\tau)$$e^{i\omega_{0}t}$+${B_{}}^*(\tau)$$e^{-i\omega_{0}t}$, where $B_{}$($\tau$)$^*$ is the complex conjugate of $B_{}$($\tau$). In the following, we will refer to $B_{}$($\tau$) and $B_{}$($\tau$)$^*$ simply as $B$ and $B^*$ respectively. Eq. (S5) can then be rewritten as: 
\begin{equation}
\begin{split}
\MoveEqLeft
        D_0^2v^{(1)} + \omega_0^2v^{(1)} + \\
        &2{D_0}{D_1}{B} e^{i\omega_{0}t}+2  \omega_{0}\gamma_{tot}{D_0}{B} e^{i\omega_{0}t}+\\
        &\omega_{0}^2 p (1-\beta {v}^2) sin (2\omega t)  {B} e^{i\omega_{0}t}+c.c.=0.
\end{split}
\end{equation}

In Eq. (S6), c.c. denotes the complex conjugate of all the terms in Eq. (S6) proportional to $e^{i\omega_{0}t}$. Eq.(S6) can be further simplified as: 
\begin{equation}
\begin{split}
\MoveEqLeft
       i\omega_{0}e^{i\omega_{0}t}[2\partial {B}/ \partial\tau+2\omega_{0}\gamma_{tot}{B}-\\&
       i\omega_{0}p(1-\beta {v}^2) sin (2\omega t){B}]+\\ &
       D_0^2 {v}^{(1)}+\omega_{0}^2 {v}^{(1)} + c.c. =0.
\end{split}
\end{equation}

In Eq. (S7), the terms proportional to $e^{i\omega_{0}t}$ consist of group of secular terms that would make the solution of ${v_{}}^{(1)}$ unbounded if their sum was not equal to zero \cite{calvanese_strinati_theory_2019}, which is not possible considering the nature of the problem we are solving. In other words, the expression multiplying $e^{i\omega_{0}t}$ in Eq. (S7) must be equal to zero, and this gives us the opportunity to retrieve a first-order differential equation in terms of $B$ that we can use to compute the real and imaginary parts of $B$, namely $B_R$ and $B_I$, respectively [see Eq. (S8)]:
\begin{equation}
\begin{aligned}
\MoveEqLeft
    {B}'=\frac{1}{4}{[(\omega_0p\beta)}(B^3-3BB^*{^2}) + \omega_0pB^*-\\
    &4\omega_{0}\gamma_{tot}{B}].
\end{aligned}
\end{equation}


The stability of a single \emph{MR} can be directly analyzed from Eq. (S8). In order to do so, we separate the real and imaginary part of $B$ by rewriting $B$ as $B_{R}$+ $i B_{I}$. This allows to rewrite Eq. (S8) as a system of two decoupled first-order differential equations [Eqs. (S9,S10)] in the variables $B_{R}$ and $B_{I}$:
\begin{equation}
\begin{split}
\MoveEqLeft
    \frac{\partial B_{R}}{\partial \tau} =B_{R}[p/4-{p\beta/2}({B_{R}}^2+3{B_{I}}^2)-\gamma_{tot}]\omega_{0},
    \end{split}
    \end{equation}
    \begin{equation}
        \begin{split}
    \frac{\partial B_{I}}{\partial \tau} =B_{I}[-p/4+{p\beta/2}({B_{I}}^2+3{B_{R}}^2)-\gamma_{tot}]\omega_{0}.
\end{split}
\end{equation}

From Eqs. (S9-S10), we can now compute the Jacobian matrix relative to the system of equations in Eqs. (S9-S10) as:

\begin{equation}
\begin{split}
    [J] = 
\begin{pmatrix}
    J_{11} & J_{12} \\
    J_{21} & J_{22}
\end{pmatrix},
\end{split}
\end{equation}
where:
\begin{align*}
J_{11} = (p/4-{3p\beta |B|^2/2}-\gamma_{tot})\omega_{0},
\end{align*}
\\[-3\baselineskip]
\begin{align*}
    J_{12} = {-3p\omega_{0}\beta B_{I} B_R},
\end{align*}
\\[-3\baselineskip]
\begin{align*}
    J_{21} = {3p\omega_{0}\beta B_{R} B_I},
\end{align*}
\\[-3\baselineskip]
\begin{align*}
   J_{22} = (-p/4+{3p\beta |B|^2/2} -\gamma_{tot})\omega_{0}.
\end{align*}


Finally, we can study the stability of the trivial solution by extracting the eigenvalues ($\lambda_{\pm}$) of $[J]$ after linearizing it around ($B_R$=0, $B_I$=0). $\lambda_{\pm}$ are found to be:
\begin{equation}
\begin{split}
\MoveEqLeft
    \lambda_{\pm}=(-4\gamma_{tot}\pm p)\omega_{0}/4 .
\end{split}
\end{equation}

From {Eq. (S12)} it is straightforward to find that $\lambda_{+}$ becomes equal to zero for $p$=$p_{th}$=$4\gamma_{tot}$, marking the transition to a nontrivial period-doubling regime.

\section{\label{sec:level1} Section S2}
\begin{center}
\textbf{Modelling a System of Coupled PFDs}
\end{center}

In this section, we study the interacting dynamics of coupled \emph{MR}s. From Fig. 1-b, we can apply {Kirchhoff's Current Law} (KCL) at the central node of the \textit{i}th-\emph{MR}, giving:

\begin{equation}
\begin{split}
\MoveEqLeft
       i_{MR}^{(i)}=i_{R_{L}}^{(i)}+\sum\limits_{j \neq i}i_{C}^{(i,j)},
\end{split}
\end{equation}

Assuming that the coupling conductances are significantly smaller than 1/$R_{L}$, the final term on the right-hand side of Eq. (S13) can be regarded as the accumulation of small fractions of $I_{MR}^{(i)}$ that flow towards the other \emph{MR}s connected to the \textit{i}th-\emph{MR}. In this scenario, $\sum\limits_{j \neq i}i_{C}^{(i,j)}$ can be simplified as:

\begin{equation}
\begin{split}
\MoveEqLeft
  \sum\limits_{j \neq i}i_{C}^{(i,j)}={R_{L}} {C_{av}}{\sum\limits_{j \neq i} \epsilon{G_{ij}}{\dot{v_j}}}.
\end{split}
\end{equation}

By using Eq. (S14) and by applying KVL to the \emph{MR} circuit shown in Fig. 1-b, we get:

\begin{equation}
\begin{split}
\MoveEqLeft
    {L} {C(t)} {\ddot{v_i}}+ v_{i}+\epsilon{R_{tot}} {C_{av}} \dot{v_i} + \\
    &{R_{L}}^2 {C_{av}}{\sum\limits_{j \neq i} \epsilon{G_{ij}}{\dot{v_j}}} =0.
\end{split}
\end{equation}

By recalling that ${\omega^2_{out}(t)}$ equals $1/[L{C(t)}]$, Eq. (S15) can be further simplified as:

\begin{equation}
\begin{split}
\MoveEqLeft
       {\ddot{v_i}}+ \omega_{out}^2 \epsilon{R_{tot}} {C_{av}} \dot{v_i}+ \omega_{out}^2 {v_i}+  \\
       &\omega_{out}^2 {R_{L}}^2 {C_{av}}{\sum\limits_{j \neq i} \epsilon{G_{ij}}{\dot{v_j}}}=0.
\end{split}
\end{equation}

By setting $\gamma_{tot}$ and $\gamma_{L}$ equal to $\omega_{out}$$C_{av}$$R_{tot}$/2 and $\omega_{out}$$C_{av}$$R_{L}$/2 respectively, rewriting $\omega_{out}^2$ using Eq. (2) and neglecting the higher order terms of $\epsilon$ once again,  we can rewrite Eq. (S16) as: 
\begin{equation}
\begin{split}
\MoveEqLeft
      {\ddot{v_i}} +  
      \omega_0^2 [1+\epsilon p (1-\beta {v_i}^2) sin (2\omega_0 t)]v_i + \\
      & \omega_0\epsilon 2\gamma_{tot}\dot{v_i} + {\omega_{0}} 2\gamma_L{R_{L}} {\sum\limits_{j \neq i} \epsilon{G_{ij}}{\dot{v_j}}}=0.
\end{split}
\end{equation}

Like in the previous section, we treat $\epsilon$ as a small expansion parameter and we consider only the terms that are at the order of $\epsilon$. For the dynamics of $v_{i}$, we first separate the quickly varying timescale from the slowly varying one, thereby expressing $v_{i}$ as ${v_{i}}^{(0)}$+$\epsilon$${v_{i}}^{(1)}$, where ${v_{i}}^{(0)}$ and  ${v_{i}}^{(1)}$ are the zeroth-order and first-order expansions of $v_{i}$, respectively. Also, following what we did for the analysis of the single \emph{MR}, the time derivatives can be rewritten as $d/dt$=$D_{0}$+$\epsilon$$D_{1}$, and $d^2/dt^2$=${D_{0}}^2$+2$\epsilon$$D_{0}$ $D_{1}$, where $D_{0}$=$\partial$/$\partial t$ and $D_{1}$=$\partial$/$\partial \tau$. This allows to rewrite Eq. (S17) as:

\begin{equation}
\begin{split}
\MoveEqLeft
      D_0^2 {v_i}^{(1)}+ \omega_0^2v_i^{(1)} + 2D_0D_1v_i^{(0)} + \\
      & 2\omega_0\gamma_{tot}D_0v_i^{(0)} + \omega_0^2p(1-\beta v_i^2)sin(2\omega t)v_i^{(0)} + \\
      & 2\omega_0\gamma_LR_L {\sum\limits_{j \neq i} D_0G_{ij} v_j^{(0)}} = 0.
\end{split}
\end{equation}

Due to the nature of the problem we are analyzing, we expect the lowest-order response of $v_i$ to vary at a rate equal to $\omega_0$. As a result, $v_i^{(0)}$ can be rewritten as $B_{i}(\tau)$$e^{i\omega_{0}t}$+${B_{i}}^*(\tau)$$e^{-i\omega_{0}t}$, where ${B_{i}}^*(\tau)$ is the complex conjugate of $B_{i}(\tau)$. This allows to rewrite {Eq. (S18)} as: 
\begin{equation}
\begin{split}
\MoveEqLeft
      D_0^2 {v_i}^{(1)}+ \omega_0^2v_i^{(1)} + 2D_0D_1B_ie^{i\omega_{0}t} + \\
      & 2\omega_0\gamma_{tot}D_0B_ie^{i\omega_{0}t} +\\
      &  \omega_0^2p(1-\beta v_i^2)sin(2\omega t)B_ie^{i\omega_{0}t} + \\
      & 2\omega_0\gamma_LR_L {\sum\limits_{j \neq i} D_0G_{ij} B_je^{i\omega_{0}t}} + c.c. = 0,
\end{split}
\end{equation}

where for simplicity we are writing $B(\tau)$ and $B(\tau)^*$ as $B$ and $B^*$ respectively. In {Eq. (S19)}, we are lumping the terms proportional to $e^{-i\omega_{0}t}$, which are the complex-conjugate of the terms proportional to $e^{i\omega_{0}t}$, into the term ``c.c.". {Eq. (S19)} can then be rewritten as:
\begin{equation}
\begin{split}
\MoveEqLeft
       i\omega_{0}e^{i\omega_{0}t}[2\partial {B_i}/ \partial\tau + 2\omega_{0}\gamma_{tot}{B_i}-\\
       &i\omega_{0}p(1-\beta {v_i}^2) sin (2\omega t){B_i}+ 2\omega_{0}\gamma_{L} {R_{L}}{\sum\limits_{j \neq i} {G_{ij}}}{B_j}] + \\
       &D_0^2 {v_i}^{(1)}+{\omega_{0}}^2 {v_i}^{(1)} + c.c. =0.
\end{split}
\end{equation}

Just like in the single-\emph{MR} case, the term multiplied by $e^{i\omega_{0}t}$ must be equal to zero. This enables us to derive a first-order differential equation that can be used to calculate the real and imaginary components of $B_i$ [see Eq. (S20)]. From Eq. (S20), we obtain the necessary means to compute the solution for any targeted combinatorial optimization problem by determining the steady-state phase value ($\Phi$) for all the \emph{MR}s used to map it:
\begin{equation}
\begin{aligned}
\MoveEqLeft
    {B_i}'(\tau)=\frac{1}{4}{[(\omega_0p\beta)(B_i^3-3}B_iB_i^{*^2}) + \omega_0pB_i^*- \\
       &4\omega_{0}\gamma_{tot}{B_i}-4\omega_{0}\gamma_{L}R_{L}{\sum\limits_{j \neq i} {G_{ij}}}{B_j}].
\end{aligned}
\end{equation}

{To model noise in our system, we mapped the Brownian thermally generated white noise voltages produced by each resistor as a Wiener process in our system of \emph{MR}-equations. It is worth emphasizing that the same approach has been used to capture the effect of noise during the analysis of SHIL IMs through the Kuramoto model} \cite{wang_oscillator-based_2017,chou_analog_2019}.


\section{\label{sec:level1} Section S3}
\subsection{\label{sec:level2} Section S3.1}
\begin{center}
\textbf{PFDs' Design}
\end{center}


Since the $PW_{spin}$ value of a PFD IM closely corresponds to the $P_{th}$ value of the constituent PFDs, it is crucial to design the PFDs in a manner that permits the attainment of the lowest possible $P_{th}$ value. In order to do so, following \cite{hussein_systematic_2020}, we must select the PFDs' components such that i) the series of $L_2$, $C_2$, $L_3$, and $C_3$ series-resonate at $\omega_{0}$; ii) the series of $L_1$, $C_1$, $L_3$, and $C_3$ series-resonate at $\omega_{in}$; iii) $L_1$ and $C_1$ behave as a band-stop filter at $\omega_{0}$; and iv) $L_2$ and $C_2$ behave as a band-stop filter at $\omega_{in}$. Starting from these design conditions, we designed the PFDs (see Fig. S1-a) we assembled in this work through a numeric optimization routine run in a commercial circuit simulator. Through this design step, we identified the following on-the-shelf components: $C_3$: Model n.SMV1236-079LF (26.75pF, tuning range = 36\%), $L_1$: Model n.1812LS334XLJC (330µH), $L_2$: Model n.1812LS-474XLJC (470µH), $L_3$:Model n.1812LS-334XLJC (330µH), $C_1$: Model n.GRM1555C1H750JA01 (75pF), and $C_2$: Model n.GRM1555C1R70WA01 (0.7pF). 

\subsection{\label{sec:level2} Section S3.2}

\begin{center}
    \textbf{Circuit Simulation of PFDs}
\end{center}

Most commercial circuit simulators struggle to detect parametric instabilities, leading to difficulties in designing parametric circuits with optimal performance \cite{hussein_systematic_2020}. Various simulation approaches have been proposed, each with its own challenges and shortcomings \cite{hussein_systematic_2020}. In particular, time-domain simulations hardly converge in the presence of points of marginal stability, like the Hopf bifurcations exhibited by parametric circuits during the activation of a period-doubling regime. In contrast, frequency-domain techniques like Harmonic Balance \cite{suarez_analysis_2009} (HB) are frequently used due to their fast computation time, even when analyzing complex nonlinear circuits. However, commercial HB simulators are unable to detect subharmonic oscillations since they assume that only the input signal's frequency or its harmonics can be generated by any circuits, thus being inherently unable to detect parametrically generated subharmonic oscillations. The power auxiliary generator (\emph{pAG}) technique has only recently been introduced to overcome this limit \cite{cassella_low_2015}. This technique forces HB simulators to consider also the frequencies that are subharmonic of the input signals' frequencies, enabling the detection of subharmonic oscillations. A \emph{pAG} is a power generator operating at $\omega_{out}$, characterized by an internal resistive impedance, $R_{AG}$. The generator is connected to the circuit on behalf of the circuit's load, $R_{L}$, and $R_{AG}$ is set to be equal to $R_{L}$. Meanwhile, the generator is configured to deliver a power level that is comparable to the noise level in the circuit, ensuring that the \emph{pAG} does not alter the operating point of any nonlinear components (see Fig. S1-a). At the same time, since the \emph{pAG} is inserted on behalf of $R_{L}$ and $R_{AG}$ is equal to $R_{L}$, the introduction of the \emph{pAG} in the circuit does not perturb the impedance seen by the nonlinear components at any frequency. This makes sure that the nonlinear dynamics of the circuit with the \emph{pAG} match exactly those of the original circuit with $R_{L}$.
\begin{figure}[t]
\includegraphics[width=\linewidth]{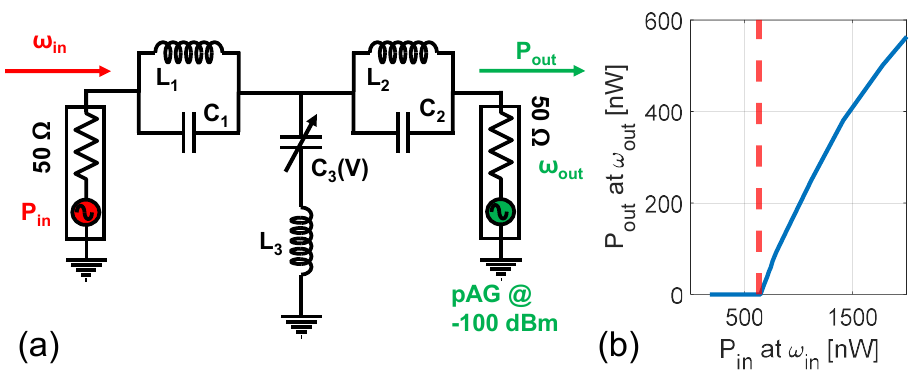}
\caption{\label{fig:epsart} a) Circuit schematic of a PFD when implementing the power auxiliary generator technique for detection of subharmonic instabilities and for evaluation of the circuit's response for input power levels exceeding the power threshold; b) Simulated $P_{out}$ vs. $P_{in}$ curve for the PFD employed in the PFD IM we assembled in this work. }
\end{figure}
The adoption of the \emph{pAG} technique in this work allowed us to determine the PFDs' optimal components, and to visualize the PFDs' electrical response for input power levels higher than $P_{th}$. This was easily done by calculating the power delivered to $R_{AG}$. In this regard, we report in {Fig. S1-b} the simulated $P_{out}$ vs. $P_{in}$ trend at $\omega_{out}$ for the PFDs we assembled in this work. Evidently, this trend shows the presence of a supercritial bifurcation for $P_{in}$ approximately equal to 600 nW, above which a significant $P_{out}$ is delivered to $R_{AG}$. Thanks to the fact that our selected annealing schedule requires driving each PFD with a power level that exceeds $P_{th}$ by a mere $0.5\%$, the $PW_{spin}$ value of PFD IMs can be can be approximated as the $P_{th}$ value of a single PFD.

\subsection{\label{sec:level2} Section S3.3}

\begin{center}
    \textbf{PFDs as Parametrons}
\end{center}

{The utilization of parametrons was proposed several decades ago as an approach to analog computing owing to their inherent phase bistability. Yet, over the last two decades circuit designers working on electronic IMs have not investigated the possibility of using parametrons to build compact electronic-based IMs able to solve large-scale COPs. This is motivated by the fact that all parametron designs reported to date required a significant power to trigger their subharmonic oscillation, even approaching the Watts-range in some cases. Also, designing parametrons in circuit simulations could not be easily done in the available circuit simulators due to constraints already discussed in Section S3.2. 
Despite the fact that PFDs have recently gathered significant interest for signal processing, frequency generation, and sensing, they can operate as ultralow-power parametrons in the framework of analog computing. In fact, PFDs can trigger parametric oscillations at exceptionally low power levels by relying on four different resonances}\cite{hussein_systematic_2020} {rather than on only one, as occurs instead in prior parametron designs. This allows to reconstruct the dynamics of an \emph{MR} while ensuring, at the same time, the highest possible efficiency in modulating the MR's resonance frequency.} {Specifically, relying on different resonant conditions, as discussed in} \cite{hussein_systematic_2020}{, permits to maximize the voltage across the nonlinear capacitor at the pump frequency for any applied input power. This allows to surpass, by orders of magnitude, the power that any other parametron reported to date has been able to demonstrate} \cite{english_an_2022,heugel_ising_2022}. {In this regard, the parametrons in} \cite{english_an_2022,heugel_ising_2022} {have $P_{th}$-values fundamentally limited by the number of resonances leveraged to reinforce the parametric oscillation conditions. For instance, when considering a quality factor for their inductors matching that of the inductors used in our assembled PFDs and when assuming the same nonlinear capacitor device, the parametron designs reported in} \cite{english_an_2022,heugel_ising_2022} {exhibit a minimum power threshold five orders of magnitude higher than what is achieved by the PFDs in this work.  This has been confirmed by circuit simulations (see Fig. S2). Note that the threshold power value reported in Fig. S2 for the paper in} \cite{heugel_ising_2022} {matches closely what was measured experimentally, considering that the authors of} \cite{heugel_ising_2022} {needed around 2.3 V for their pump voltage to activate the subharmonic oscillation, which corresponds to more than 100 mW when using a conventional 50 ohm signal generator}.

\begin{figure}[t]
\includegraphics[width=\linewidth]{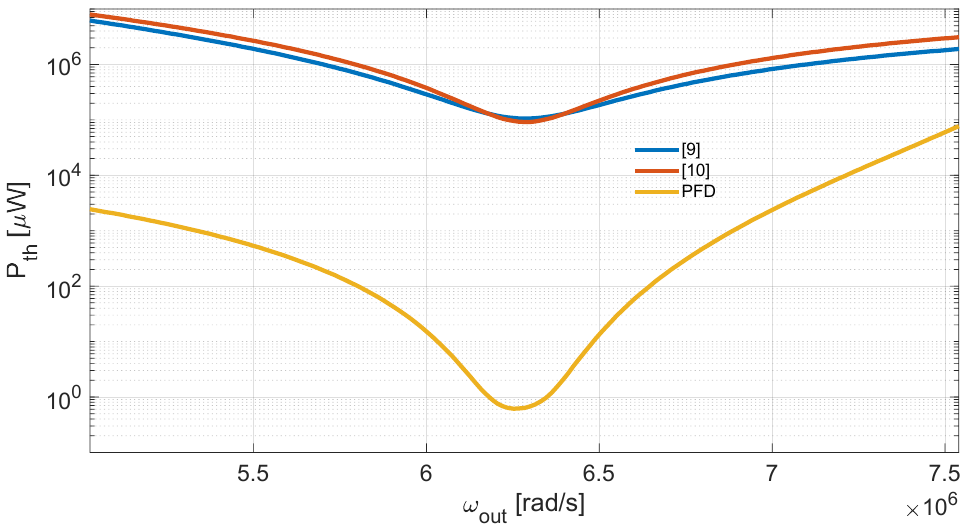}
\caption{\label{fig:epsart} Simulated trends of the optimal power threshold vs. the output natural frequency for the reported PFD design and for the parametron designs described in \cite{english_an_2022,heugel_ising_2022}. These trends have been computed by assuming the same input signal, with a frequency matching what has been used in this work for the assembled PFD IMs.}
\end{figure}

\section{\label{sec:level1} Section S4}

\begin{center}
    \textbf{Coupled PFDs as Coupled Parametrons}
\end{center}

{In order to understand what drives the activation of out-of-phase or in-phase signals in a network of $MR$s, it is necessary to recall the key dynamical feature governing the activation of parametric oscillations in a circuit containing a nonlinear capacitor. Subharmonic oscillations are triggered in such a circuit when the modulation of the nonlinear capacitor's reactance generates enough parametric gain to compensate for the losses in the circuit, which in our case include the $MR$'s intrinsic losses as well as the losses generated by the equivalent electrical resistance connected to it. By extending this concept to the illustrative system of two resistively coupled $MR$s (see Fig. S3-a), it becomes evident that the two possible non-trivial solutions corresponding to in-phase and out-of-phase output signals are distinguished by different power thresholds.}


\begin{figure}[b]
\includegraphics[width=\linewidth]{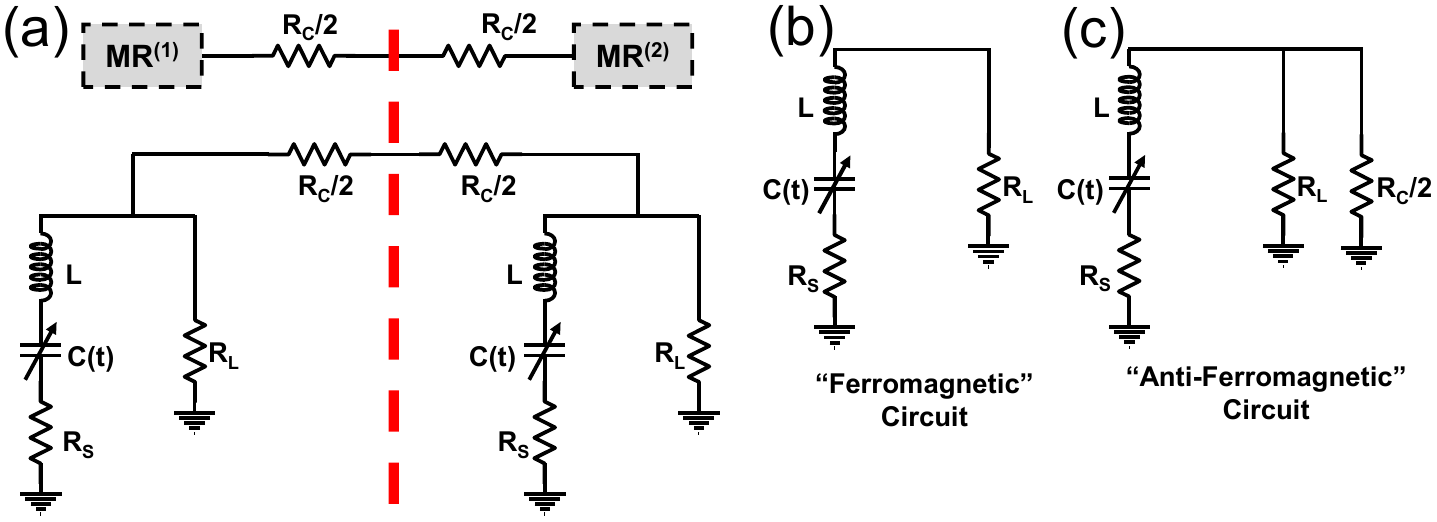}
\caption{\label{fig:epsart} a) System-level and circuit-level schematic of two resistively coupled $MR$s; b)-c) Equivalent ``ferromagnetic" and ``anti-ferromagnetic" circuits for the two coupled $MR$s system in (a).}
\end{figure}

{In this context, if the two $MR$s in Fig. S3-a were to settle into a state where their output signals are in-phase, no current would flow through the coupling resistor ($R_C$). As a result, the total losses that each $MR$ would have to compensate for activating its subharmonic oscillation would be determined by the sum of $R_S$ and $R_L$. Differently, if the two $MR$s in Fig. S3-a were to settle to a state corresponding to out-of-phase output signals, there would be current flowing into $R_C$. In other words, an out-of-phase coupling between the two $MR$s would result in each $MR$ being connected to an equivalent resistance equal to the parallel combination of $R_L$ and $R_C$/2} \cite{rober_novel_2013}, {which is obviously lower than either resistors. These simple considerations suggest that the dissipative coupling for the circuit in Fig. S3-a leads to a lower power threshold for activating out-of-phase subharmonic oscillations than in-phase $MR$s' output signals. As a result, when the pump power is activated and increased, the system of two $MR$s will reach the power threshold for the activation of out-of-phase parametric oscillations first. In other words, the resistively coupled $MR$s in Fig. S3-a will always exhibit out-of-phase output signals when considering realistic $R_L$ and $R_C$ values.}

{The conclusion we have just drawn based on a simple observation of the circuit in Fig. S3-a can be also formalized and validated by using a circuit simulation approach. In particular, one can analyze the two coupled $MR$s in Fig. S3-a by studying the response of two separate circuits (labeled as the ``ferromagnetic" and ``anti-ferromagnetic" circuits in Fig. S3-b,c) including only one $MR$ and different electrical terminations ($R_L$ for the ferromagnetic circuit and the parallel of $R_L$ and $R_C$/2 for the anti-ferromagnetic circuit). }
{We can retrieve the power threshold for each circuit by using numerical methods (Fig. S4-a), clearly proving that the anti-ferromagnetic circuit has a lower power threshold than the ferromagnetic circuit. 
This has been also confirmed for the numerical case analyzed in Fig. S4-b by running a circuit simulation of the full circuit formed by two $MR$s (Fig. S3-a). As evident from Fig. S4-b, our simulation confirms that the two $MR$s show indeed out-of-phase output signals. 

It is worth noting that a ferromagnetic coupling between two PFDs can be also implemented if needed to reconstruct an in-phase relationship between the PFDs' output signals. This can be done by using coupling capacitors. In fact, relying on capacitive components permits to detune the resonance frequency of the PFDs' output mesh when considering the anti-ferromagnetic circuit. This detuning causes an increase of the power threshold of the anti-ferromagnetic circuit, while not affecting the power threshold of the ferromagnetic circuit.}
{Having explained what rules the phase-relationship between the output signals of two $MR$s, it is easier to understand what is the principle of computation for PFD IMs {for solving the polynomial-time M\"obius Ladder problems investigated in the Main Manuscript} \cite{english_an_2022,wang_bifurcation_2023}. In this regard, {when solving polynomial-time problems with a ground-state solution that can be mapped to the eigenvalues of the coupling matrix}, PFD IMs operate under the principle that the lowest energy solution identifying the ground-state corresponds to the combination of spin-orientations that requires the lowest power to be activated} \cite{wang_coherent_2013,mohseni_ising_2022,kalinin_computational_2022,wang_bifurcation_2023}. {However, for general computational purposes, we iterate that PFD IMs obey a Lyapunov function describing both amplitude and phase dynamics} \cite{roychowdhury_global_2022}.


\begin{figure}[b]
\includegraphics[width=\linewidth]{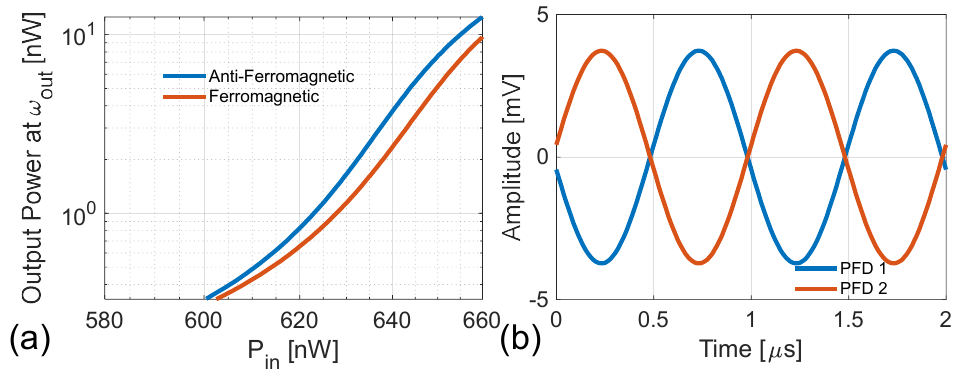}
\caption{\label{fig:epsart} a) Numerical simulations of the power dissipated by $R_L$ at $\omega_{out}$ for both the ferromagnetic (in orange) and the anti-ferromagnetic (in blue) circuits shown in Fig. S3-b,c. Evidently, the anti-ferromagnetic circuit exhibits a lower $P_{th}$ than the ferromagnetic circuit; b) Simulated output voltage waveforms across $R_L$ for each of the two $MR$s coupled by $R_C$.}
\end{figure}


\section{\label{sec:level1} Section S5}

\begin{center}
    \textbf{Impact of the Annealing Schedule on the Performance of PFD IMs}
\end{center}

As also demonstrated by others \cite{calvanese_strinati_can_2021, chou_analog_2019}, the ability of IMs to achieve an accurate solution for a given combinatorial optimization problem decays rapidly as the number of unknowns in the problem ($N$) increases. Clearly, this poses significant challenges in achieving IMs that can accurately solve realistic, large-scale problems where $N$ is much higher than $10^3$ \cite{cen_large-scale_2022}. To address this limitation in OIMs, others have proposed the use of an annealing schedule. By incorporating an annealing schedule, in fact, the effectiveness of OIMs in preventing convergence to local minima and successfully identifying an accurate solution may be greatly enhanced \cite{bohm_order_2021}, thereby leading to higher probabilities-of-success. However, as others have shown for SHIL IMs \cite{bybee_efficient_2022}, relying on an annealing schedule also leads to longer times-to-solution, creating a trade-off between solution accuracy and energy-to-solution that is difficult to overcome \cite{mohseni_ising_2022}. It is then crucial to examine how using an annealing schedule impacts on the performance of PFD IMs. Therefore, during the analytical evaluation of the computing performance of PFD IMs, we employed an exponential-based annealing schedule [see Eq. (S22)]. Our purpose was to gradually increase the value of $p$ from 0.995$p_{th}$ (e.g., right below threshold) to 1.005$p_{th}$ (e.g., right above threshold), with a rate set by $\tau_{ann}$.

\begin{equation}
    p(t) = p_{th}[0.995+0.01(1-e^{-t/\tau_{ann}})]
\end{equation}

\begin{figure}[b]
\includegraphics[width=\linewidth]{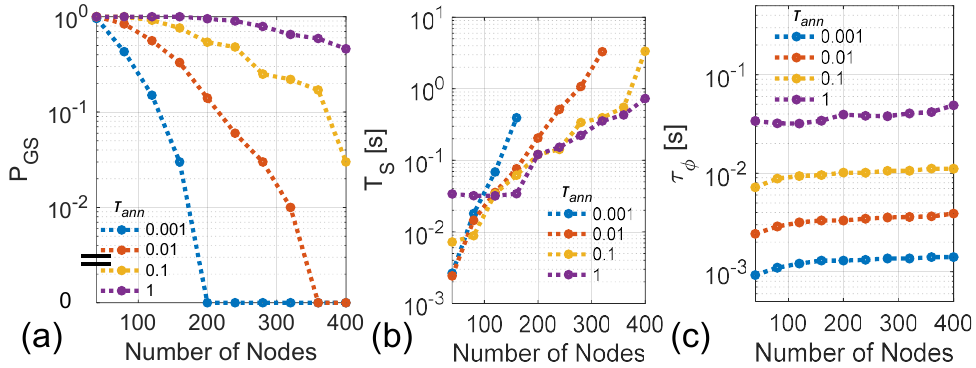}
\caption{\label{fig:epsart} Numerically computed trends vs. $N$ of a) $P_{GS}$, b) $T_S$, and c) $\tau_{\phi}$ when considering \emph{Q} = 50 and when assuming different values (in seconds) for $\tau_{ann}$. {All points are averaged over 100 runs.} }
\end{figure}


\begin{figure}[t]
\includegraphics[width=\linewidth]{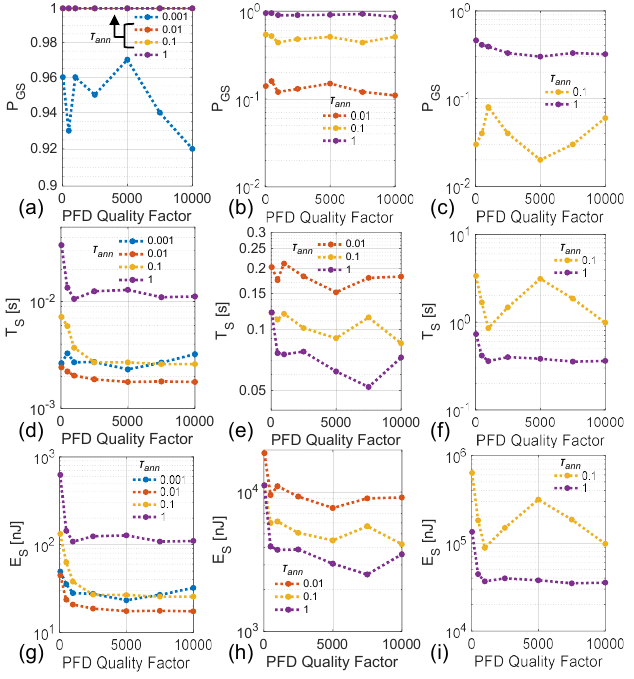}
\caption{\label{fig:epsart} Numerically computed trends vs. Q of a) $P_{GS}$ when $N$ = 40, b)$P_{GS}$ when $N$ = 200, c) $P_{GS}$ when $N$ = 400, d) $T_S$ when $N$ = 40, e) $T_S$ when $N$ = 200, f) $T_S$ when $N$ = 400, g) $E_S$ when $N$ = 40, h) $E_S$ when $N$ = 200, and i) $E_S$ when $N$ = 400 for different values (in seconds) of $\tau_{ann}$. {Note that the $\tau_{ann}$ = 0.001 sec annealing profile was omitted from subsections b), e), and h) because its simulated $P_{GS}$ values were 0\% for all values of $Q$ except for $Q$ = 1000 ($P_{GS}$ = 1\%) and 7500 ($P_{GS}$ = 1\%). This same annealing schedule was also excluded from any sub-figure corresponding to $N$ = 400, or c), f), and i), since the resultant $P_{GS}$ was always 0\%. Similarly, the annealing schedule of $\tau_{ann}$ = 0.01 sec was omitted from subsections c), f), and i) because its calculated $P_{GS}$ was mostly 0\% with the exception of when $Q$ = 2500 ($P_{GS}$ = 1\%). All points are averaged over 100 runs.} }
\end{figure}

We assessed the performance of PFD IMs  for different $\tau_{ann}$ values and for $N$ varying from 40 to 400.{ Our findings reveal that employing larger $\tau_{ann}$ values results in higher $P_{GS}$ values when solving large size M\"obius ladder problems. As such, for large problem sizes, the adoption of slower annealing schedules can also shorten $T_S$ and, consequently, reduce $E_S$. In contrary, for small problem sizes, a high $P_{GS}$ is attained without requiring a long annealing schedule, meaning that for small problem sizes, $T_S$ and $E_S$ can be degraded by the adoption of a high $\tau_{ann}$ value (see Fig. S5).}

The advantage of employing our annealing schedule to enhance the overall performance of PFD IMs has also been investigated for different $Q$ values, ranging from 50 to 10000, and for various $N$ values, ranging from 40 to 400 (see Fig. S6). {Varying \emph{Q} does not alter significantly $P_{GS}$, especially when using the larger $\tau_{ann}$ values required to solve large size problems. On the other hand, using resonators with \emph{Q} higher than 500 in the PFDs' design leads to improved $T_S$ and $E_S$, especially for larger values of $\tau_{ann}$.} Meanwhile, as discussed in the main manuscript, using higher-$Q$ resonators in the PFDs' design enables lower $PW_{spin}$, which is crucial to ensure an accurate resolution of large-scale combinatorial optimization problems while still consuming a low power.

{To elucidate on the origin of the computing performance improvement due to the adoption of an annealing schedule, we investigated the impact of $\tau_{ann}$ on the convergence dynamics of the system. In this regard, as $p$ is gradually increased through annealing from a value below $p_{th}$ to one above $p_{th}$, we found that our system experiences a reduction in amplitude heterogeneity }\cite{leleu_destabilization_2019,albash_analog_2019,nifle_new_1992}. {It is well established that in oscillator-based IMs, cases exhibiting substantial amplitude heterogeneity are more prone to converging towards energy minima levels that no longer correspond to the ground state of the initial Ising Hamiltonian mapping the problem to be solved} \cite{albash_analog_2019}. {In this regard, differences in steady-state amplitudes among the \emph{MRs} alter the nature of the problem that is solved by effectively changing the coupling strengths among the \emph{MRs} with respect to the what is set by $G_c$ and by the problem graph} \cite{albash_analog_2019}. {It is through the introduction of an annealing schedule that we are able to guide the system towards a more accurate solution of the problem by enforcing that the selected coupling weights between nodes remain nearly unchanged throughout the duration of the system's phase synchronization. To analytically verify this finding, we conducted a study where we analyzed the average coefficient of variation ($CV$, representing the ratio of the standard deviation and the mean) of the steady-state amplitudes of all the oscillators used to solve an $N$ = 400 M\"{o}bius ladder problem. For each set of 100 runs of this problem, we considered different $\tau_{ann}$ values matching those used in Fig. S6. The extracted distributions of $CV$ and $P_{GS}$ vs. $\tau_{ann}$ can be found in Fig. S7. Evidently, as the annealing gets slower (corresponding to an increase in the value of $\tau_{ann}$), $CV$ reduces and the system becomes more likely to minimize the Ising Hamiltonian of the specified original problem.} {}

\begin{figure}[t]
\centering
\includegraphics[width=\linewidth]{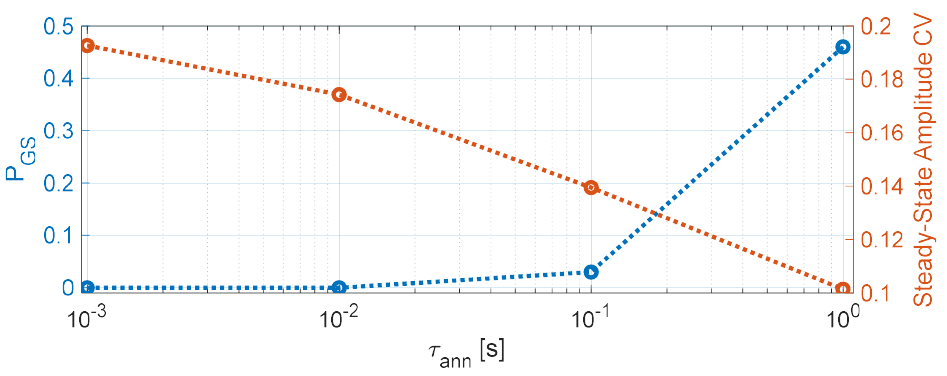}
\caption{\label{fig:wide} Numerically computed trends vs. $\tau_{ann}$ of $P_{GS}$ and $CV$ for a 400-node M\"{o}bius ladder problem.}
\end{figure}

\begin{figure}[b]
\includegraphics[width=\linewidth]{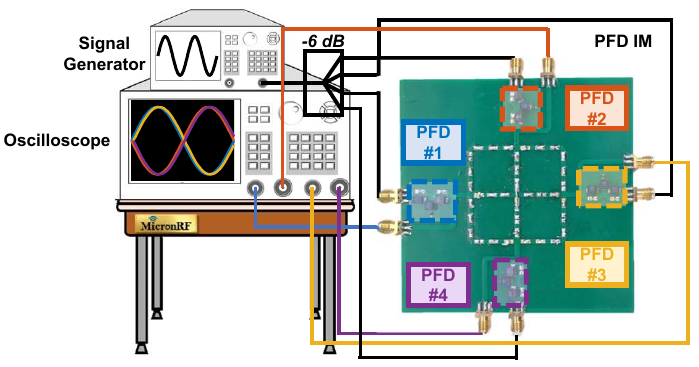}
\caption{\label{fig:epsart} Experimental setup we used for measuring the output waveforms of our assembled 4-node PFD IM. Each PFD is driven by the same power-divided input signal, and each PFD's output voltage is measured by using an oscilloscope. }
\end{figure}

\section{\label{sec:level1} Section S6}
\begin{center}
\textbf{Experimental Setup}
\end{center}

The PFD IM assembled in this work included four identical PFDs, together with the routing lines and the soldering pads required for programming the PFDs' coupling (see Fig. S8). The PFDs' output ports were coupled by {2$k\Omega$} resistors, following the graph of the specific problem to solve. In this regard, the absence of a resistor between two PFDs represented a disconnection between them in the problem graph. All the PFDs were driven by the same 2 MHz input signal, equally split. Due to the reduced problem-size we considered for the validation of our PFD IM, we did not implement an annealing step. In fact, as shown in Fig. S5-a, executing an annealing step is only necessary for large $N$-values \cite{chou_analog_2019}.

In order to validate the proper functioning of our PFD IM, we connected the PFDs' output ports to various ports of an oscilloscope (Model No. InfiniiVision DSOX6004A) and we measured their time-domain output voltage. Also, we arbitrarily assigned PFD 1 (Fig. S8) as our reference PFD for the evaluation of the computed solution and, consequently, for the extraction of the computed maximum number of cuts as described in the main manuscript. In this regard, an approach relying on determining the cross-correlation between any given PFD's output signal and the reference's output signal was employed to determine the phase-shift between each PFD's output signal and the output signal of PFD 1. This allowed us to determine whether the solution computed by our PFD IM was correct. Moreover, for each problem we tackled, we used our PFD IM to search for the solution five times and the system converged to the correct solution of the problem graph during each run. Finally, in order to switch between different problems, we manually changed the set of coupling resistors connected to the circuit.  

Fig. S9 shows all the investigated Max-Cut problem graphs and their solutions, along with the PFDs' measured output waveforms. It is worth mentioning that our PFD IM retrieved the correct solution for all problems we attempted and for all the problem runs we executed. A description of the correct solution for each problem we considered is available in \cite{chou_analog_2019}.

\begin{figure}[t]
\includegraphics[width=\linewidth]{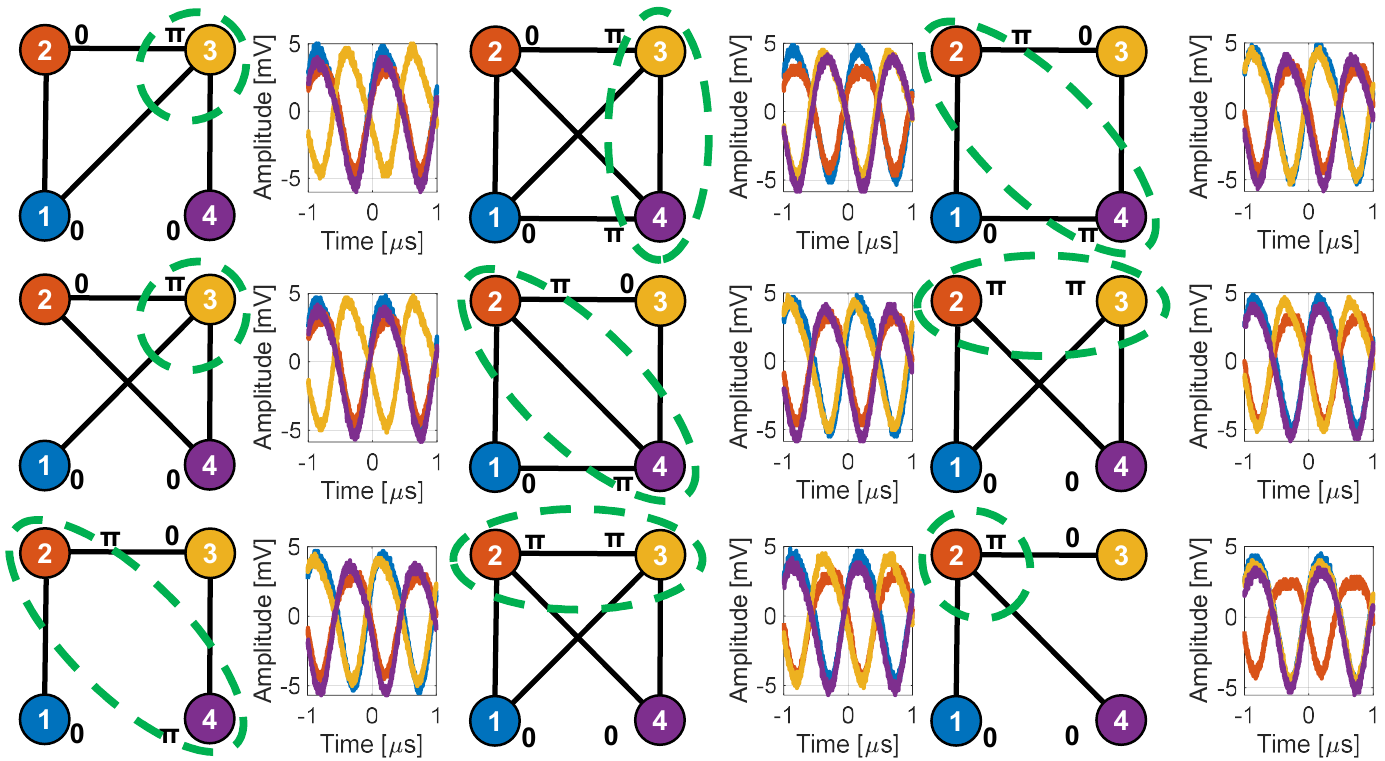}
\caption{\label{fig:epsart} Graphs and PFDs’ output voltage waveforms relative to all nine Max-Cut problems attempted and correctly solved with the PFD IM built in this work. }
\end{figure}

\bibliography{TeX/PRL_2023_Paper_updated}